\documentclass[sigconf]{acmart}

\usepackage{subcaption} 
\usepackage{tabularx}
\usepackage{adjustbox}
\usepackage{amsthm}
\usepackage{multirow}
\usepackage{enumitem}

\newcommand{\hl}[1]{#1}

\newcommand{\nop}[1]{}
\theoremstyle{definition}
\newtheorem{mydef}{Definition}
\newcommand{\argmax}{\arg\,\max} 
\newcommand{\ignore}[1]{}

\newcommand\blfootnote[1]{%
  \begingroup
  \renewcommand\thefootnote{}\footnote{#1}%
  \addtocounter{footnote}{-1}%
  \endgroup
}

\AtBeginDocument{%
  \providecommand\BibTeX{{%
    \normalfont B\kern-0.5em{\scshape i\kern-0.25em b}\kern-0.8em\TeX}}}

\setcopyright{acmcopyright}
\copyrightyear{2020}
\acmYear{2020}
\acmDOI{10.1145/1122445.1122456}

\acmConference[SIGIR '20]{SIGIR '20: 43rd International ACM SIGIR Conference on Research and Development in Information Retrieval}{July 25--30, 2020}{Xi'an, China}
\acmBooktitle{SIGIR '20: 43rd International ACM SIGIR Conference on Research and Development in Information Retrieval,
  July 25--30, 2020, Xi'an, China}
\acmPrice{15.00}
\acmISBN{978-1-4503-XXXX-X/18/06}

\begin{document}

\title{Learning to Ask Screening Questions for Job Postings}


\author{Baoxu Shi$^\star$ \quad Shan Li$^\star$ \quad Jaewon Yang \quad Mustafa Emre Kazdagli \quad Qi He}
\affiliation{%
  \institution{LinkedIn}
}
\email{{dashi,shali,jeyang,ekazdagli,qhe}@linkedin.com}







\renewcommand{\shortauthors}{Baoxu Shi, Shan Li, Jaewon Yang, Mustafa Emre Kazdagli, and Qi He}

\begin{abstract}
At LinkedIn, we want to create economic opportunity for everyone in the global workforce. A critical aspect of this goal is matching jobs with qualified applicants. To improve hiring efficiency and reduce the need to manually screening each applicant, we develop a new product where recruiters can ask screening questions online so that they can filter qualified candidates easily. To add screening questions to all $20$M active jobs at LinkedIn, we propose a new task that aims to automatically generate screening questions for a given job posting. To solve the task of generating screening questions, we develop a two-stage deep learning model called Job2Questions, where we apply a deep learning model to detect intent from the text description, and then rank the detected intents by their importance based on other contextual features. Since this is a new product with no historical data, we employ deep transfer learning to train complex models with limited training data. We launched the screening question product and our AI models to LinkedIn users and observed significant impact in the job marketplace. During our online A/B test, we observed $+53.10\%$ screening question suggestion acceptance rate, $+22.17\%$ job coverage, $+190\%$ recruiter-applicant interaction, and $+11$ Net Promoter Score. In sum, the deployed Job2Questions model helps recruiters to find qualified applicants and job seekers to find jobs they are qualified for.


\blfootnote{$^\star$ These authors contributed equally.}

\end{abstract}


\ignore{

\ccsdesc[500]{Computer systems organization~Embedded systems}
\ccsdesc[300]{Computer systems organization~Redundancy}
\ccsdesc{Computer systems organization~Robotics}
\ccsdesc[100]{Networks~Network reliability}
}
\keywords{screening question generation, applicant screening}


\maketitle

\section{Introduction}

\begin{figure}[t] 
\centering
  \begin{subfigure}[b]{\linewidth}
    \centering
    \includegraphics[width=.98\textwidth]{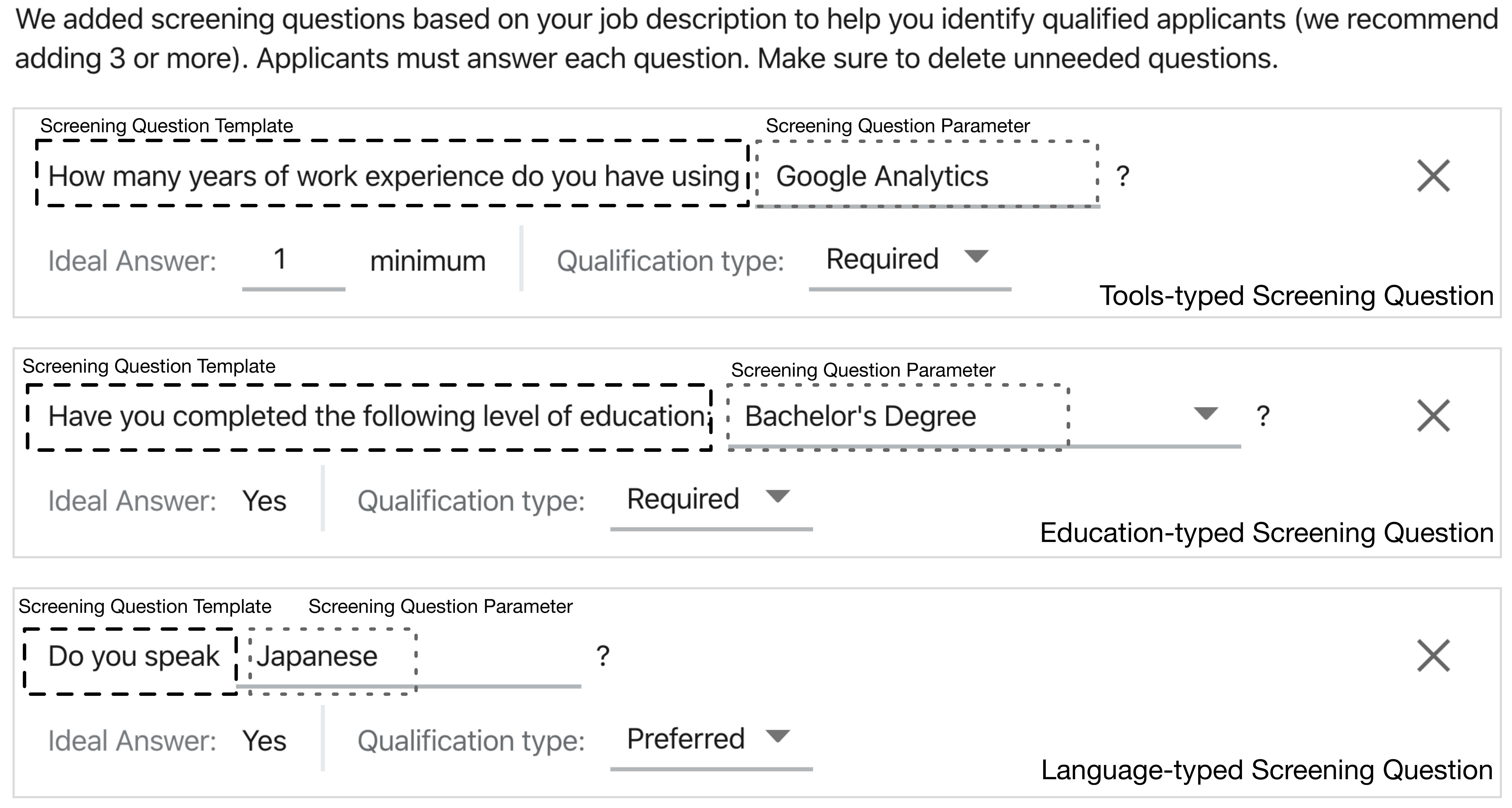}
  \end{subfigure}%
  \caption{Screenshot of screening questions suggested to a job posting being posted on LinkedIn.}
\vspace{-.5cm}
\label{fig:sq_screenshot} 
\end{figure}

LinkedIn is the largest hiring marketplace in the world, hosting over $20$ million active job postings that are created across various channels, including LinkedIn's on-site recruiting products and integrations with external hiring products.

In hiring, interviewing applicants is costly and inefficient. Therefore, recruiters typically screen the applicants in the pool by their profile and conduct additional phone screenings before sending out interview invites. According to our user research study, approximately $70\%$ of phone screenings end up finding out the applicant is missing basic qualifications such as work authorization/visa, minimum years of experience, or degree requirements. Also, the majority of the applications for coveted jobs disappear in the hiring funnel because recruiters do not have time to review them all.

To address such hiring inefficiency, researchers have proposed models to ease the manual workload by estimating person-job fit automatically. Existing methods aim to match job postings based on the members' experiences~\cite{qin2018enhancing} or based on members' profile attributes~\cite{ha2016search,diaby2013toward,zhu2018person,le2019towards}. However, these models heavily rely on the assumption that applicants' online profile and resume are always up-to-date and contain all the information that hiring companies need. As we show in Sec.~\ref{sec:case_studies}, the member profile is not the perfect source for modeling applicants because 1) members do not update their profiles promptly, and 2) there is often a gap between what members present in their profile and what employers want to know. Moreover, in Sec.~\ref{sec:case_studies} we also find that job posting text is sub-optimal for modeling job qualifications due to trivial and unnecessary requirements.

Based on the above observations, we decide to design a new Screening Question (SQ) -based online screening product for LinkedIn to assess job applicants automatically. To be specific, we proactively ask job-specific questions as shown in Fig.~\ref{fig:sq_screenshot} to applicants and assess them using the answers they provide when applying for the job. Compared to member profiles, the answers collected by SQs are most recent and contain all the facts employers want to learn.


There are two significant product challenges for designing a successful SQ-based online screening product. Firstly, the product should provide an easy way to add SQs to jobs. If we ask recruiters to reformulate their job postings into SQs manually, such an excessive need for human efforts will forbid us to add SQs to all $20$ million jobs on LinkedIn. Secondly, SQs should help recruiters identify qualified applicants quickly. If we use unstructured text questions~\cite{du_learning_2017,chali2018automatic} to present SQs, one job requirement may have many different expressions and hence hard for recruiters or AI models to interpret the intent of the SQ, group SQs with same intent together, and categorize applicants based on their answers to SQs.

In this work, we address the above product challenges by designing and productionizing a Screening Question Generation (SQG) model called Job2Questions, which automatically generates structured SQs for a given job posting. By developing a machine learning SQG model, we no longer rely on human input and can apply the SQG model to generate SQs for all $20$M jobs on LinkedIn.
By generating structured SQs in the format of (\textit{template}, \textit{parameter}) as illustrated in Fig.~\ref{fig:sq_screenshot}, SQs will have a unified internal representation that describes SQs' intent (template) and focus (parameter) precisely. To ensure SQ quality, we asked hiring experts to design and review the SQ templates and corresponding parameter lists.
Using structured representation instead unstructured text avoids SQ ambiguity and discrepancy across different jobs. This also makes it easier for AI models and recruiters to group and screen candidates based on specific intent such as education, language, and others.

Although researchers have studied the Question Generation (QG) task extensively, SQG cannot be viewed as a simple application of QG methods because it poses many unique challenges as follows.



\noindent{}\textbf{Diversified input styles and topics.} Unlike QG datasets which are often shorter passages focusing on a few specific topics, the input of SQG are lengthy text having both different narrative styles across different industries and also various topics ranging from company introduction, requirements, to benefits. As shown in Tab.~\ref{tab:qg-dataset}, the average number of words and sentences per LinkedIn job posting is larger than other common QG datasets. We believe a good SQG model needs to be able to process long text and general enough to handle job postings from different industries.

\noindent{}\textbf{Job marketplace domain-specific.} Majority of the QG methods~\cite{tang2017question,zhou_neural_2018,gao2019difficulty} are designed to generate questions to test the cognitive skills of readers~\cite{chen_learningq_2018}. To generate QGs that represent important job qualifications, a good SQG model needs to be domain-specific and have deep understanding of the job marketplace. In fact, generic QG models yield embarrassing results for the SQG task. Given the text of a \textsf{Staff Software Engineer} job posting, QG methods return \textit{What is to enable others to derive near-limitless insights from LinkedIn's data?}~\cite{heilman_question_2009} or \textit{what does Experience stand for?}~\cite{du_learning_2017}, which are not SQs as they do not represent job qualifications.

\noindent{}\textbf{Low online inference latency.} Lastly, QG models are designed without explicit latency constraints. As shown in Tab.~\ref{tab:qg-time}, QG methods usually have $20+$ms latency. However, SQG model has a more strict latency requirement because recruiters expect SQG model to provide screening questions right after they entered the job description. To avoid sluggish performance and poor customer experience, a good SQG model needs to have a simple yet effective architecture in order to keep the inference latency within an acceptable range.



%

With all above challenges in mind, here we propose a two-step SQG model named Job2Questions that given the content of a job posting, first generates all possible structured SQ candidates using a deep learning model, and then ranks and identifies top-$k$ screening questions as the model output. 

In candidate generation, we divide job postings into sentences and generate all SQ candidates by converting each SQ-eligible sentence to structured (\textit{template}, \textit{parameter}) pairs. To get the template of the sentence, we solve multiclass classification in which one sentence is classified into one of the predefined templates. The challenge is to develop a deep, fast model that can understand the semantic meaning of the job posting text with a small number of labeled examples. We apply deep transfer learning~\cite{chung_supervised_2018} with Deep Averaging Network~\cite{iyyer_deep_2015} to achieve both speed and accuracy. In terms of parameter entities, we used an in-house entity linking system to tag out mentions in the sentence and link them to the corresponding entities. For question ranking, we build an XGBoost pairwise ranking model to sort screening questions using extensive job and question features. 



The contributions of this work are summarized as follows:

\begin{itemize}[leftmargin=*]
    \item To the best of our knowledge, this is the first work on the Screening Question Generation (SQG) task, which generates structured screening questions to help assess job applicants.
    \item We proposed and deployed the first SQG model, Job2Questions, to production to help millions of jobs finding qualified applicants and help hundreds of millions of members to identify qualified jobs.
    \item During offline evaluation, the proposed Job2Questions model improved the AUROC of both template classification and question ranking by $178\%$ and $27.4\%$, respectively.
    
    \item Job2Questions significantly improved the online SQ suggestion quality by $+53.10\%$ acceptance rate and $+22.17\%$ job coverage. Jobs adopted SQ suggestions yielded $190\%$ more recruiter-applicant interactions. These improvements increase the Net Promoter Score~\cite{reichheld2003one} by $11$ points for recruiters who use Job2Questions.
    
    \item We conducted extensive analyses of the Job2Questions results and obtained exciting insights about the quality of member profile, requirements mentioned in the job posting, and different applicant screening focuses across the job marketplace.
\end{itemize}

\begin{table}[t]
    \caption{Statistics of popular question generation datasets.}
    \label{tab:qg-dataset}
    \begin{adjustbox}{max width=\linewidth}
    \begin{tabular}{l | c c}
    Dataset & Avg. words/doc & Avg. sents/doc \\
    \midrule
    SQuAD~\cite{rajpurkar_squad_2016} & $135$ & $5$ \\
    RACE~\cite{lai_race_2017} & $323$ & $18$ \\
    \midrule
    LinkedIn & $\mathbf{584}$ & $\mathbf{40}$
    \end{tabular}
    \end{adjustbox}
\end{table}

\begin{table}[t]
    \caption{Empirical CPU inference time per sentence. J2Q-TC-DAN is our current in production model.}
    \label{tab:qg-time}
    \begin{adjustbox}{max width=\linewidth}
    \begin{tabular}{c | c c | c c c}
            & Rule-based & Seq2Seq & & Our SQG Models & \\
    \midrule
    Latency & H\&S~\cite{heilman_question_2009} & NQG~
    \cite{du_learning_2017} & BOW-XGB & \textbf{J2Q-TC-DAN} & J2Q-TC-BERT \\
    \midrule
       (ms) & $24$ms & $62$ms & $4$ms & $\mathbf{9}$ms & $100$ms 
    \end{tabular}
    \end{adjustbox}
\end{table}

\section{Related Work}

\begin{figure*}[t] 
\centering
  \begin{subfigure}[b]{\linewidth}
    \centering
    \includegraphics[width=.95\textwidth]{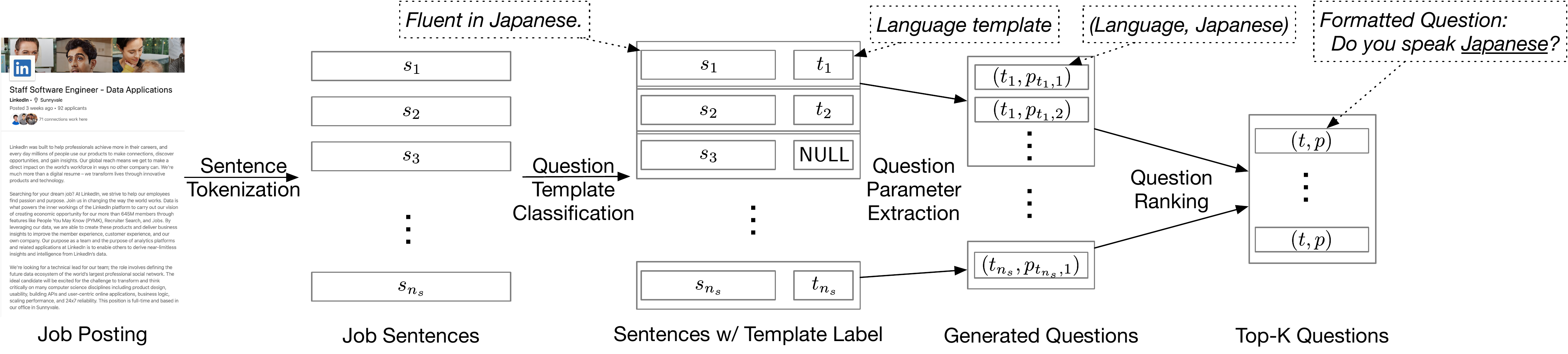}
  \end{subfigure}%
  \caption{Overview of the Screening Question Generation task and its sub-tasks.}
\label{fig:sqg_overview} 
\end{figure*}

\noindent{}\textbf{Rule-based Question Generation}. Rule-based models usually transform and formulate the questions based on the text input using a series of hand-crafted rules. ELIZA~\cite{weizenbaum1966eliza} generates question responses for conversations using human-made, keyword-based rules. Mitkov uses language patterns and WordNet to create multiple-choice questions~\cite{mitkov2003computer}. Heilman introduced a rule-based model to generate reading comprehension questions~\cite{heilman_question_2009}. Despite the high precision, these rule-based models are not scalable and require immense human efforts and domain-specific expertise, which is not suitable for our case where it is hard to conclude patterns for jobs from different industries.

\noindent{}\textbf{Context-only Neural Network Question Generation}. Recently, neural network models that generate questions based on the given context have shown promising results with little human intervention. Du et al.~\cite{du_learning_2017} and Chali et al.~\cite{chali2018automatic} employed Seq2Seq model with attention mechanism for question generation tasks and trained the model in an end-to-end fashion. These methods are not designed for generating screening questions from job posting text for applicant assessment, which is different from traditional reading comprehension question generation tasks. Moreover, the unstructured, freeform questions generated by Seq2Seq models often do not have clear-cut-intent and too vague for job posters and downstream machine learning models to interpret and categorize. Lastly, the latency of Seq2seq models is also less ideal for fast online inference.

\noindent{}\textbf{Answer-aware Neural Network Question Generation}. To simplify the question generation task and handle cases where one input sentence maps to multiple questions, researchers proposed answer-aware question generation models where both the context and the answer of the target question is known. Tang et al.~\cite{tang2017question} improved the Seq2Seq model with a global attention mechanism that leverages additional answer information in the postprocessing step by replacing out-of-vocabulary words with the word in the answer having the highest relevance score among the attention distribution. Zhou et al.~\cite{zhou_neural_2018} proposed a Seq2Seq model that takes both the context and answer as input and generates questions using answer words-copying~\cite{gulcehre2016pointing}. Sun et al.~\cite{sun2018answer} further empower the QG model by explicitly leveraging the answer embedding and modeling the distance between the answer and the context. Gao, et al.~\cite{gao2019difficulty} extend the traditional QG problem to Difficulty-controllable Question Generation (DQG) in which questions were generated based on the difficulty level designated given one pair of the context and the answer. However, the answer is not available for the screening question generation task, which means these methods are not applicable to our case. 

\noindent{}\textbf{Person-Job fit}. Unlike recommender systems~\cite{paparrizos_machine_2011,kenthapadi_personalized_2017,huang_online_2019} that recommend items by predicting member actions, our work is more related to person-job fit, which aims at identifying if an applicant is qualified for the job.
Recently, DuerQuiz~\cite{qin2019duerquiz} is proposed to create in-depth skill assessment questions to test if the applicant is good at certain hard skills such as Machine Learning. But it does not consider other general qualifications that are crucial for applicant screening, such as education background, work authorization, year-of-experience, or soft skills. APJFNN~\cite{qin2018enhancing} is developed to predict person-job fit by comparing the job description and applicants’ work experience in their resume. APJFNN does not have good explainability because it does not attribute the decision to a single requirement. Other resume-based methods~\cite{zhu2018person,le2019towards} may return sub-optimal assessment if certain information is missing in the resume. 
In general, none of these person-job fit models have studied the task of modeling person-job fit by generating explicit screening questions from job postings.

\section{Problem Statement}\label{sec:problem_statement}
The Screening Question Generation (SQG for short) task aims to generate screening questions from the job posting text. Job applicants will provide their answers to these questions during the job application. Based on the answers the applicants provide, the recruiters will identify qualified candidates. LinkedIn will also recommend other jobs to the applicants by matching their answers to the screening questions of other jobs and identify the ones that the applicants are qualified for.

At LinkedIn, we decide to use structured screening questions in the form of (\textit{template}, \textit{parameter}) instead of freeform text. For example, we use (\textit{How many years of work experience do you have using}, \textit{Java}), with the first part of the pair being template and the second part being parameter. We do it for the following reasons:
\begin{itemize}[leftmargin=*]
    \item \textbf{Structured questions ensure question quality}. By predefining the question types and possible parameters, we can ensure the screening question is unambiguous and reduce the chance of introducing inappropriate questions;
    \item \textbf{Structured questions have clear intent}. Unlike freeform text questions, the intent of structured questions are strictly defined by the question template. Therefore, job posters can easily group and screen candidates based on certain intent, e.g. education background, experience in multiple industries, or the list of tools they are familiar with. 
    \item \textbf{Structured questions standardize questions across jobs}. By limiting screening questions to have pre-defined templates and parameters, questions from different jobs will have exactly the same representation. This property makes it possible for us to recommend jobs that the applicants may be qualified for by comparing their answers to other jobs’ screening questions.
\end{itemize}

Based on the above three reasons, here we define the SQG task as inferring structured screening rather than free form questions from job posting text.

\begin{mydef}\label{def:job_sqg}
\textbf{Screening Question Generation.} Given the text of a job posting $j=\{w_1, \ldots, w_{n_w}\}$, where $w$ represents words in the job and $n_w$ denotes the total number of words in the job. \textbf{Screening Question Generation (SQG)} returns $k$ top-ranked structured screening questions $\{(t, p)| t \in \mathbf{T}, p \in \mathbf{P}_t\}$, where $\mathbf{T}$ is a set of pre-defined templates, and $\mathbf{P}_t$ is a set of pre-defined parameters used by template $t$.
\end{mydef}

For example, given a job posting of \textsf{Staff Software Engineer - Data Applications} posted by LinkedIn, the Screening Question Generation model should return a list of screening questions in the format of template and parameter pairs such as (\textit{Have you completed the following level of education}, \textit{Bachelor’s Degree}) and (\textit{How many years of work experience do you have using}, \textit{Java}).

However, designing a SQG model that can generate screening questions using the whole job posting as input is challenging. The job postings are longer and much noisier compared to the SQuAD~\cite{rajpurkar_squad_2016} dataset, where each passage is relatively short and only focuses on one topic. As we shown in Tab.~\ref{tab:qg-dataset}, the average length of a job posting is four times longer than the SQuAD passages. Moreover, job postings usually cover a wide range of topics, including company description, job functions, benefits, compensation, schedules, disclaimers, and many others that are not related to the screening question generation process.

Inspired by the fact in Question Generation (QG for short) tasks that the majority ($99.73\%$~\cite{du_learning_2017}) of the questions could be derived from a single sentence, we hypothesize that the SQG problem can also be modeled by a sentence-level model. 

Based on the above assumption, we propose a four-component sentence-level SQG framework and illustrate it in Fig.~\ref{fig:sqg_overview}. As shown in Fig.~\ref{fig:sqg_overview}, we first tokenize the given job posting into sentences, and then we run a question template classification model to detect the most probable template for each sentence. For every sentence that has a valid, non-$\mathsf{NULL}$ template, we first use the template-dependent parameter extractor to extract possible parameters, and then construct a list of screening question candidates using the extracted parameters and the template. Lastly, we aggregate all question candidates and use a question ranking model to pick $k$ top-ranked template-parameter pairs as the final suggested screening questions for the given job posting. Next, we will provide the formal definition of the sentence-level SQG task and its sub-tasks.

\begin{mydef}\label{def:sent_sqg}
\textbf{Sentence-level Screening Question Generation}. Given job posting $j=\{s_1, \ldots, s_{n_s}\}$, \textbf{Sentence-level Screening Question Generation} extracts a list of screening question candidates $\mathbf{Q}_j = \bigcup_{s_i\in j}\{(t, p) | t=\textsf{TC}(s_i), p\in \textsf{PE}(s_i, t)\}$, where $\textsf{TC}$ is some question template classification model, and $\textsf{PE}$ is some parameter extraction model, and output $k$ top-ranked structured screening questions $\{(t, p)| t \in \mathbf{T}, p \in \mathbf{P}_t\}=\textsf{QR}(\mathbf{Q}_j, j, k)$, where $\textsf{QR}$ is some question ranking model.
\end{mydef}

In Def.~\ref{def:sent_sqg}, we enforce a one-to-one mapping between the sentence and the question template. Based on our observation, we find that the ratio of sentences that map to only one question is $88.9\%$. The rest $11.1\%$ sentences, on the other hand, usually maps to multiple questions with the same template but different parameters. For example, ``\textit{4+ years experience programming experience in Java and C/C++}'' can be converted into two screening questions with the same template (\textit{How many years of work experience do you have using}, \textit{Java}) and (\textit{How many years of work experience do you have using}, \textit{C/C++}).

Next, we will formally define three sub-tasks of the sentence-level SQG task, namely question template classification (TC), template parameter extraction (PE), and question ranking (QR).

\begin{mydef}\label{def:tc}
\textbf{Question Template Classification}. Given a sentence $s = \{w_1, \ldots, w_{n_w}\}$ from job posting $j$ where $w$ are word tokens, \textbf{Question Template Classification} (TC) predicts the question template $t_s \in T$ of $s$ or $\textsf{NULL}$ if $s$ does not match any template.
\end{mydef}

\begin{mydef}\label{def:pe}
\textbf{Template Parameter Extraction}. Given a sentence $s = \{w_1, \ldots, w_{n_w}\}$ from job posting $j$ and predicted template $t_s$, \textbf{Template Parameter Extraction} (PE) extracts a list of possible parameter values $\mathbf{P}_{s,t_s}$ from $s$ with respect to template $t_s$.
\end{mydef}

Note that for a given sentence $s$, Def.~\ref{def:tc} may return \textsf{NULL} if $s$ should not be converted into any screening question. Note that SQG is different from the traditional QG settings where the input passage always maps to one or more questions~\cite{duan_question_2017,zhao_paragraph-level_2018}. In SQG, a large portion of the sentences in the job posting is irrelevant to the qualification evaluation of an applicant, and therefore should not be converted into screening questions.

After getting the screening question candidate set $\mathbf{Q}_j$, we use a question ranking model to rank all the questions and return the top-k as the generated screening questions of job posting $j$.

\begin{mydef}\label{def:qr}
\textbf{Question Ranking}. Given a list of screening question candidates $\mathbf{Q}_j$, \textbf{Question Ranking} (QR) ranks them into an ordered list based on $\mathsf{Pr}(\mathsf{accepted}|(j,t,p))$, the probability that job posters will add screening question $(t,p)$ to job $j$.
\end{mydef}

In the following sections, we will describe the data collection strategy and the model design of our proposed sentence-level SQG model, Job2Questions.

\section{Data Preparation}
In this section, we will describe two methods we used, namely crowdsourcing and user feedback, to collect training and high-quality evaluation data for the template classification and question ranking tasks. Note that we leverage our existing, in-house entity linking system as the parameter extractor, therefore we will omit the data preparation for the parameter extraction component in this section. The statistics of the two datasets we collected in this section is described in Tab.~\ref{tab:sqg-stat}.

\begin{table}[t]
    \caption{Screening Question Generation dataset statistics.}
    \vspace{-.2cm}
    \label{tab:sqg-stat}
    \begin{tabular}{l | r r}
    Task & Train & Test \\
    \midrule
    Question Template Classification & $7,053$ & $3,648$ \\
    Question Ranking & $88,354$ & $22,055$ \\
    \end{tabular}
    \vspace{-.2cm}
\end{table}

\subsection{Question Template Classification}
Given a sentence $s$, the first step of SQG is to predict its question template $t$. In order to train a template classification model, we need to collect sentence-template $(s, t)$ pairs. 

We performed a crowdsourcing task to collect labeled sentence-template pairs. Tab.~\ref{tab:cs-example} shows two examples of the crowdsourcing annotation task we designed to collect labeled data. The sentences and screening questions are generated as follows: we first recognize entities from all sentences, and collect a list of sentences that contain valid parameter entities; then for each sentence-parameter $(s, p)$ pair, we generate a screening question $(t, p)$ for $s$, where $p$ can be used as $t$'s parameter ($p \in P_t$); lastly, we randomly sample a subset of these generated $(s, t, p)$ triples, convert them into the format shown in Tab.~\ref{tab:cs-example}, and ask human annotators to label these sentence-question pairs. We consider the sentence-template pair $(s, t)$ pair is positive if the human labeler labels at least one triple from $\{(s, (t, p))|p\in\mathbf{P}_t\}$ as directly related. Otherwise we consider that sentence $s$ maps to \textsf{NULL} template $(s, \textsf{NULL})$.

\begin{table}[t]
    \caption{Examples of the crowd sourcing annotation task.}
    \label{tab:cs-example}
    \begin{adjustbox}{max width=\linewidth}
    \begin{tabularx}{0.9\textwidth}{X}
    \toprule
    \textbf{Is the given sentence from job description directly related to the given screening question?}\\
    \midrule
    \textbf{Sentence from job description}: \\
    \quad\quad Post-graduate or PhD in Computer Science or Machine Learning related degree with a focus on NLP;\\
    \textbf{Screening Question}: \\
    \quad\quad Have you completed the following level of education: \underline{Ph.D.}? \\
    \\
    \toprule
    \textbf{Is the given sentence from job description directly related to the given screening question?}\\
    \midrule
    \textbf{Sentence from job description}: \\
    \quad\quad Performing annual and periodic Fair Lending and UDAAP analysis and reporting utilizing CRA Wiz and R Studio . \\
    \textbf{Screening Question}: \\
    \quad\quad How many years of work experience do you have using \underline{R}?
    \end{tabularx}
    \end{adjustbox}
    \vspace{-.2cm}
\end{table}

\subsection{Question Ranking}\label{sec:data_prep_qr}
As shown in Fig.~\ref{fig:sqg_overview}, once we have a list of screening question candidates, the next step is to rank them and pick the top-$k$ ranked questions as the output of the SQG model. The objective of question ranking is to predict the probability of a screening question $(t,p)$ been added to job posting $j$ by the job poster. To train such a ranking model, we need to collect the corresponding $(j,t,p)$ triples for model training.

Although we can ask job posters to manually add screening questions to jobs, such approach only provides positive labeled triples. The challenge for question ranking data collection is the lack of negative labeled data. The random generated negative data are easy to separate and cannot help improve the model performance. Another approach is to randomly pick auto-generated screening questions that do not match manual-added questions from job posters as negative labeled data. Such negative labeled triples may have a high false-negative rate: job posters did not add such questions manually because they simply forgot about it.

Therefore, to collect high-quality question ranking data, we need to explicitly ask job posters to provide negative labeled data. In this work, we first designed a simplified sentence-level SQG model which is described in Sec.~\ref{sec:exp_setting} as the BOW-XGB model, deployed it in production to provide screening question suggestions to job posters, and then collect the labeled question ranking triples using job posters' feedbacks. Namely if a job poster accepts a suggestion or adds a new screening question, we generate a positive labeled $(j,t,p)$ triple. If job poster rejects a screening question suggestion, we generate a negative labeled $(j,t,p)$ triple accordingly. We collected $110,409$ labeled triples and group them into two triple sets $(j,t,p)\in D^+$ and $(j,t,p) \in D^-$, where $D^+$ contains all positive-labeled data and $D^-$ is a triple set of negative-labeled data.

\section{Job2Questions}\label{sec:job2questions}
After describing the problem formulation and data preparation process for the sentence-level screening question generation (SQG) task, here we describe the detailed design of our production sentence-level SQG model, the Job2Questions model. We will describe its three core components as shown in Fig.~\ref{fig:sqg_overview}, namely the question template classification, question parameter extraction, and question ranking.

\subsection{Question Template Classification}
\begin{figure}[t] 
\centering
  \begin{subfigure}[b]{\linewidth}
    \centering
    \includegraphics[width=.95\textwidth]{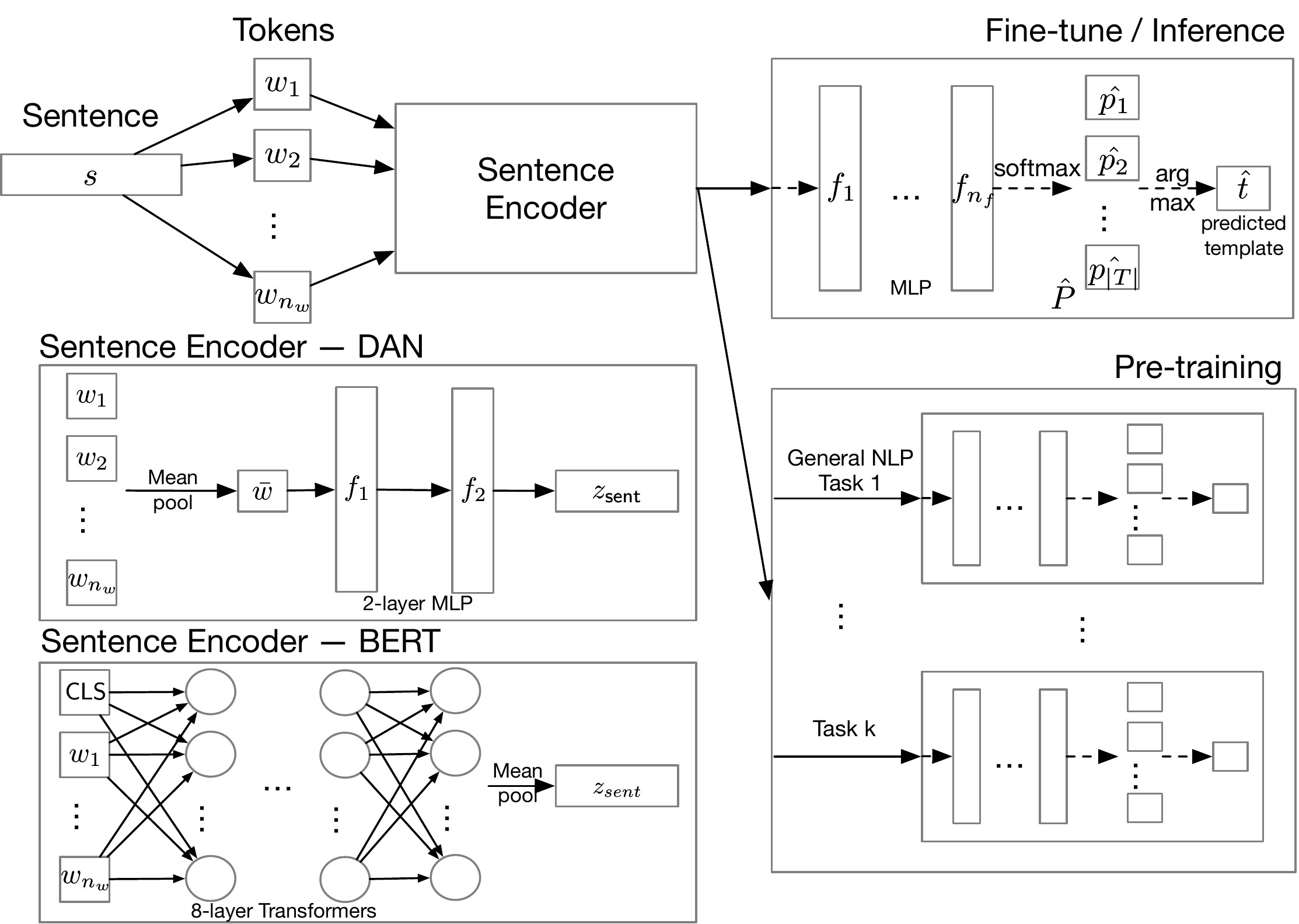}
  \end{subfigure}%
  \caption{Question Template Classification model. We first pretrain Sentence Encoders (DAN/BERT) with general NLP tasks~\cite{devlin_bert:_2018,iyyer_deep_2015}, then \hl{conduct task-specific fine-tuning.} During inference, the model takes sentence as input and use sentence encoder + task-specific MLP to predict template.}
  \vspace{-.4cm}
\label{fig:template_classification} 
\end{figure}

As the first component of the Job2Questions model, question template classification takes a raw sentence as input and predicts its most probable template label, or NULL if it is not eligible. Here in this work, we treat this task as a multiclass classification task and consider template labels and the non-eligible NULL as classes. The overview of the question template classification component is shown in Fig.~\ref{fig:template_classification}. 

As shown in Fig.~\ref{fig:template_classification}, for a given sentence $s$, we first tokenize it into \hl{word} tokens $\{w_1, \ldots, w_{n_w}\}$, and then use a sentence encoder to convert the tokens into a sentence embedding vector $z_{\textsf{sent}}$. The generated sentence embedding vector $z_{\textsf{sent}}$ is then sent to a neural network model to predict its most probable class label $\hat{t}$. 

Because the training data set (around $8$k) is relatively small compared to tens of millions of jobs posted on LinkedIn, it definitely does not contain all the words in the vocabulary and does not cover all the creative ways recruiters describe job requirements. To address this issue, we decide to utilize multi-task transfer learning~\cite{pan_survey_2010} to pre-train the sentence encoding model with multiple natural language understanding tasks, and then use transfer learning to fine-tune the trained model with our question template classification task. As shown in Fig.~\ref{fig:template_classification}, we first pre-train the sentence encoder with general tasks, and then fine-tune the sentence encoder with the task-specific MLP using our template classification data. 

Here we propose two methods to encode sentence into embeddings, a simple and fast Deep Average Network (DAN)~\cite{iyyer_deep_2015} model and a more advanced Deep Bidirectional Transformers (BERT)~\cite{devlin_bert:_2018} model. We choose DAN due to its simplicity and competitive performance compared to relative computational expensive models such as CNN~\cite{jiao_convolutional_2018} and LSTM~\cite{lin_structured_2017} models. We also used BERT as our sentence encoder to see how advanced NLP model can help improve the performance in this specific task.

\hl{The DAN model first average the embedding of the input tokens into $\bar{w} = \frac{1}{n_w}\sum_i^k w_i$, and then pass it through two fully-connected layers to get the sentence representation $z_\textsf{sent}=\sigma(\sigma(\bar{w}W_1 + b_1) W_2 + b_2)$.}

BERT, on the other hand, uses Transformer layer~\cite{vaswani_attention_2017} to encode the input sentence to embedding. It is defined as follows
\small
\begin{align}
\begin{split}
    \textsf{TFLayer}(h^{n-1}) &= \textsf{FC}(\textsf{MultiAttn}(h^{n-1}));\\
    \textsf{FC}(x) &= \textsf{relu}(xW_1 + b_1)W_2 + b_2;\\
    \textsf{MultiAttn}(h^{n-1}) &= \textsf{concat}(head_1(h^{n-1}),\ldots,head_k(h^{n-1}))W^O; \\
    \textsf{head}_i(h^{n-1}) &= \textsf{softmax}\left(\frac{(h^{n-1}W_q^i)(h^{n-1}W_k^i)}{\sqrt{d_k}}\right)(h^{n-1}W_v^i).
\end{split}
\end{align}
\normalsize
\noindent{}where $h^{n-1}$ is the output of the previous Transformer layer. Here we use a BERT model with $8$ Transformer layers, and define the output sentence embedding $z_{\textsf{sent}}$ as the meanpooling result of the last transformer layer's output. For simplicity, we omit batch normalization~\cite{ioffe_batch_2015} and residual connections~\cite{he_deep_2016} in the equations.

After we obtain the sentence embedding $z_{\textsf{sent}}$, we then pass it through a multilayer perceptron network (MLP) where each fully-connected layer is defined as $f(x) = \textsf{relu}(xW + b),$
and the last layer of the MLP is defined as 
$\hat{P} = \textsf{softmax}(f(x)W + b),$
where the output $\hat{P}$ is the categorical probability distribution of each class. Finally, we pick the most probable class $\argmax(\hat{P})$ as the final predicted template label. To train the model, we use a binary-cross entropy loss 
\small
\begin{equation}
    \mathcal{L}(P,\hat{P}) = -\sum_{i}p^i\log\hat{p}^i,
\end{equation}
\normalsize
\noindent{where} $P$ is the ground truth, $p^i$ and $\hat{p}^i$ are the ground truth and predicted probability of $i^{\textsf{th}}$ template respectively. We use the Adam optimizer~
\cite{kingma_adam_2017} to optimize the model parameters.

\subsection{Question Parameter Extraction}

Given a job posting, we first break it down into sentences and get each sentence's template label using the above question template classification model. For sentences that have a valid non-NULL template and require a parameter, we call our in-house entity linking system to detect the corresponding template parameter by tagging specific types of entities from the given sentence in real-time. Note this work is not about designing the entity linking system so we will only give a brief overview of the system below.

To find template parameters, the system first tag possible entity mentions from sentences. We utilize an in-house, comprehensive entity taxonomy that contains large sets of entity surface forms to identify possible entity mentions from the given text. In our current production model, we support four types of entities, namely education degrees, tool-typed skills, spoken languages, and credentials (certifications and licenses).


After we identified entity mentions from job sentences, we then use a feature-based regression model to link the mention to an entity in the taxonomy. Besides global features such as mention frequency, we also employed many contextual features such as POS tag, context n-grams, and the cosine similarity between the FastText~\cite{joulin_bag_2016} embeddings of the mention and its context. These contextual features help our model to identify invalid mentions such as \textsf{Bachelor's degree} in ``\textit{We provide bachelor party supplies}'' or \textsf{Chinese} language in ``\textit{Our clients include European and Chinese companies}''. Here we choose FastText instead of other methods such as LSTM~\cite{lample_neural_2016} or charCNN~\cite{ma_end--end_2016} because FastText has a lower latency and works reasonably well for identifying and linking the four entity types we listed above.

Finally, entity mentions with a confidence score that passes the given threshold will be considered as template parameters of the given sentence $s$ and template $t_s$.

\subsection{Question Ranking}\label{sec:job2questions_question_ranking}
After we get all the question candidates in the format of template label and parameter pairs $(t,p)$ from the given job posting, the next step is to rank the candidates and find questions that are helpful for the hiring process. Because determining whether or not a screening question is helpful for the hiring process is non-trivial, in this work we rely on recruiters and job posters to label what screening questions are best for the hiring process. Hence, we use the screening questions added and rejected by them as the ground truth labels and define the question ranking objective as predicting the likeliness of a job poster adds a screening question candidate $(t,p)$ to a job posting $j$. 
\small
\begin{equation}\label{eq:qr_prob}
    	\textsf{Pr}(\textsf{accepted} | j, t, p) = sigmoid(f(x_{j,t,p})),
\end{equation}
\normalsize
\noindent{where $f$ is the scoring function, $x_{j, t, p}$ is the feature vector with respect to the given job $j$, template label $t$, and parameter $p$.}

The features we used to construct $x_{j,t,p}$ can be group into three groups, job-side features, question-side features and job-question interactive features.

\noindent{}\textbf{Job-side features}: Job attributes such as job's title, industry, company, location, and others. We use $27$ different features to represent jobs.

\noindent{}\textbf{Question-side features}: Screening question attributes such as question template type, parameter value, template classification score, and entity linking system's confidence score. We use $5$ features to represent questions.

\noindent{}\textbf{Job-Question interactive features}: We generate interactive features by computing the Pointwise Mutual Information (PMI) between job- and question-side features. The PMI is defined as follows:

\small
\begin{equation}\label{eq:qr_interaction_feature_pmi}
    	\textsf{PMI}(F_j ; F_q) = log\frac{\textsf{Pr}(F_j,F_q)}{\textsf{Pr}(F_j)\textsf{Pr}(F_q)},
\end{equation}
\normalsize

\noindent{}where $F_j$ and $F_q$ are the job- and question-side categorical features respectively. Here we use PMI value to quantify the discrepancy between the probability of correspondence of a job-side and a question-side event given both joint and individual distribution. In total we use $135$ interactive features in our question ranking model.

\begin{table*}[ht]
    \caption{Question template classification offline evaluation.}
    \label{tab:tc-eval}
    \centering
    \begin{adjustbox}{max width=\linewidth}
        \begin{tabular}{l|c|cccccccccccccc}
         &  & \multicolumn{2}{c}{NULL} & \multicolumn{2}{c}{Work Auth} & \multicolumn{2}{c}{Sponsorship} & \multicolumn{2}{c}{Education} & \multicolumn{2}{c}{Language} & \multicolumn{2}{c}{Credential} & \multicolumn{2}{c}{Tools} \\
        Model & Overall Acc. & Prec. & Rec. & Prec. & Rec. & Prec. & Rec. & Prec. & Rec. & Prec. & Rec. & Prec. & Rec. & Prec. & Rec. \\
        \midrule
        BOW-XGB & $0.3547$ & $0.1336$ & $\mathbf{0.5988}$ & $0.4118$ & $0.1386$ & $0.5000$ & $0.0051$ & $0.5644$ & $0.8519$ & $\mathbf{0.9888}$ & $0.3346$ & - & - & $0.5329$ & $0.4863$ \\
        J2Q-TC-CNN & $0.8640$ & $0.5408$ & $0.2536$ & $0.9397$ & $0.9639$ & $0.9488$ & $0.9915$ & $0.9199$ & $0.8857$ & $0.9328$ & $0.9680$ & $0.6865$ & $\mathbf{0.9227}$ & $0.9037$ & $0.8151$ \\
        J2Q-TC-NNLM & $0.8765$ & $0.6250$ & $0.2871$ & $0.9425$ & $0.9716$ & $0.9514$ & $0.9915$ & $0.8984$ & $0.9343$ & $0.9598$ & $0.9710$ & $0.7019$ & $0.9040$ & $0.9193$ & $0.8374$ \\
        J2Q-TC-DAN & $0.8798$ & $0.6330$ & $0.3301$ & $0.9333$ & $\mathbf{0.9742}$ & $0.9538$ & $0.9887$ & $0.9056$ & $0.9314$ & $0.9648$ & $0.9564$ & $0.7227$ & $0.8827$ & $0.9068$ & $0.8664$ \\
        J2Q-TC-BERT & $\mathbf{0.9138}$ & $\mathbf{0.6688}$ & $0.4928$ & $\mathbf{0.9592}$ & $0.9691$ & $\mathbf{0.9592}$ & $\mathbf{0.9944}$ & $\mathbf{0.9282}$ & $\mathbf{0.9600}$ & $0.9655$ & $\mathbf{0.9767}$ & $\mathbf{0.8564}$ & $0.8747$ & $\mathbf{0.9197}$ & $\mathbf{0.9465}$
        \end{tabular}
    \end{adjustbox}
    \vspace{-.2cm}
\end{table*}

After describing the feature vector $x_{j,t,p}$, next we present the scoring function $f$. Here we use XGBoost as the scoring function and therefore rewrite Eq.~\ref{eq:qr_prob} as 

\small
\begin{equation}
    \textsf{Pr}(\textsf{accepted} | j, t, p) = sigmoid\left(\sum_k f_k(x_{j,t,p})\right)
\end{equation}
\normalsize

\noindent{}where $f_k$ is the $k^\textsf{th}$ tree of the model. We use the following loss function to optimize the question ranking model

\small
\begin{equation}
\begin{split}
    \mathcal{L} =& -\sum_{\langle j,t,p\rangle\in\mathbf{D}^+}\log\left(\sum_{k}f_k\left(x_{j,t,p}\right)\right) \\
    &- \sum_{\langle j,t,p\rangle\in\mathbf{D}^-}\log\left(1 - \sum_{k}f_k\left(x_{j,t,p}\right)\right) + \sum_{k}\Omega(f_k),
\end{split}
\end{equation}
\normalsize

\noindent{}where $D^+$ and $D^-$ are the positive and negative $(j, t, p)$ triple sets collected using the job posters’ feedback described in Sec.~\ref{sec:data_prep_qr}. $f_k$ represents the $k^\mathsf{th}$ tree in the boosted-tree model, $\Omega(f_k) = \gamma T + \frac{1}{2}\lambda||\mathbf{w}||^2$ is the regularization term that penalizes the complexity of tree $f_k$, in which $T$ denotes the number of leaves in tree $f_k$, $\mathbf{w}$ is the leaf weights, $\gamma$ and $\lambda$ are the regularization parameters.

\section{Experiments}
In this section, we conducted extensive evaluations on the proposed Job2Questions (J2Q for short) model. The promising offline and online A/B test results demonstrate the effectiveness of the proposed Job2Questions model in terms of providing high quality screening question suggestions, helping recruiters identify qualified applicants, suggesting qualified jobs to members, and boosting recruiter-applicant interactions. The large-scale case studies on Job2Questions result also reveal many interesting insights that help us better understand the job marketplace.

\subsection{Experiment Setting}\label{sec:exp_setting}
The evaluated question template classification models are:
\begin{itemize}[leftmargin=*]
    \item \textbf{BOW-XGB}: A non-neural network baseline which tokenize the input sentence into bag-of-word vectors and then trained an XGBoost~\cite{chen_xgboost_2016} model to predict the template label.
    \item \textbf{J2Q-TC-NNLM}: J2Q template classification model which uses a simple feed-forward neural network language model (NNLM)~\cite{bengio_neural_2003} as the sentence encoder.
    \item \textbf{J2Q-TC-DAN}: J2Q template classification model with deep averaging networks~\cite{iyyer_deep_2015} as the sentence encoder.
    \item \textbf{J2Q-TC-CNN}: J2Q template classification model with CNN-based universal sentence encoder~\cite{jiao_convolutional_2018} as the sentence encoder.
    \item \textbf{J2Q-TC-BERT}: J2Q template classification model with BERT~\cite{devlin_bert:_2018} as the sentence encoder.
\end{itemize}
All above models are trained using the same dataset as described in Tab.~\ref{tab:sqg-stat}. For neural network J2Q-TC-$\star$ models, we use the public-available pre-trained models~\cite{cer_universal_2018,devlin_bert:_2018} as initialization and fine-tune them accordingly. For J2Q-TC-\{NNLM, DAN, CNN\}, we set the learning rate to $1e-3$, batch size to $256$, and drop-out rate to $0.4$. For J2Q-TC-BERT, we further truncate the input sentence to $32$ tokens and set the learning rate to 5e-5. All models are trained for at most $100$ epochs with a $3$-layer MLP.

We also evaluated the following question ranking models:

\begin{itemize}[leftmargin=*]
    \item \textbf{Rule-based}: A non-machine learning baseline. It sorts questions based on the template classification model score and re-ranks them using business rules, e.g. education and work authorization questions always rank at the top.
    \item \textbf{J2Q-QR-LR}: A logistic regression model that ranks candidates using job and question side features. 
    \item \textbf{J2Q-QR-XGB-pointwise}: The proposed question ranking model trained using a pointwise loss.
    \item \textbf{J2Q-QR-XGB-pairwise}: The proposed question ranking model trained using a pairwise loss.
\end{itemize}
The J2Q question ranking model is trained with $110,409$ labeled $(j, t, p)$ triples collected via job posters feedback. We divide the data into $70-20-10$ training, evaluation, and validation sets. We explored and anchored the XGBoost hyper-parameters as follows: number of trees is $100$, depth is $5$, $\eta$ is $0.7$, and $\gamma$ is set to $0$.


\subsection{Offline Evaluation}

\begin{figure*}[t] 
\centering
  \begin{subfigure}[b]{0.25\linewidth}
    \centering
    \includegraphics[width=.9\textwidth]{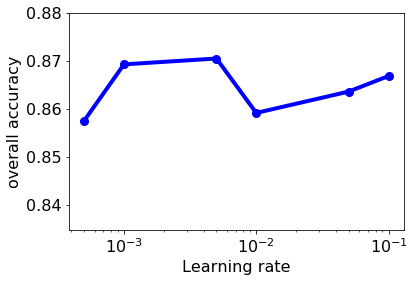}
  \end{subfigure}%
    \begin{subfigure}[b]{0.25\linewidth}
    \centering
    \includegraphics[width=.9\textwidth]{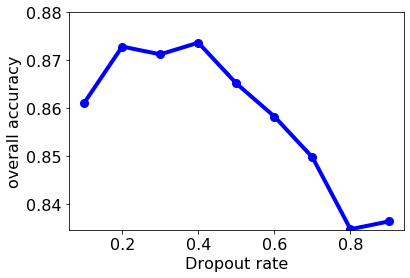}
  \end{subfigure}%
    \begin{subfigure}[b]{0.25\linewidth}
    \centering
    \includegraphics[width=.9\textwidth]{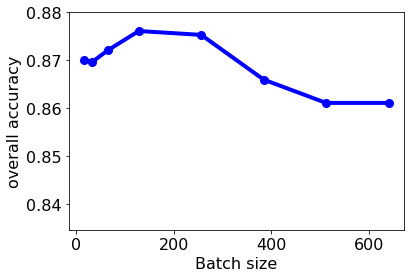}
  \end{subfigure}%
    \begin{subfigure}[b]{0.25\linewidth}
    \centering
    \includegraphics[width=.9\textwidth]{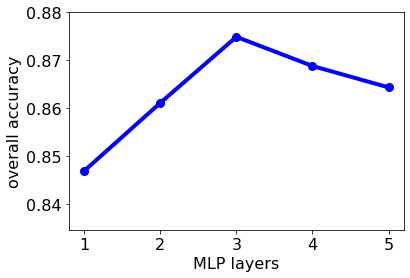}
  \end{subfigure}%
\vspace{-.2cm}
  \caption{Hyper-parameter sensitivity test for the in-production J2Q-TC-DAN model.}
 \vspace{-.2cm}
\label{fig:hyper_parameter_test} 
\end{figure*}

\subsubsection{Question Template Prediction.} We evaluate the question template classification performance on the crowdsourced dataset described in Tab.~\ref{tab:sqg-stat}, which contains $6$ different template labels and a special label NULL for sentences that cannot be convert into SQ. We report the precision/recall of each template label and the overall accuracy in Tab.~\ref{tab:tc-eval}. We found that all deep learning models outperformed the baseline by a large margin, and BERT yields the best overall accuracy, outperforming the second best DAN model by $3.9\%$. However, as shown in Tab.~\ref{tab:qg-time}, the CPU inference time of BERT ($100$ms) is $10$-times longer than the DAN model ($9$ms), making it less practical for online CPU inference. 

\vspace{-.2cm}
\subsubsection{Hyper-parameter Test.} To understand how different hyper-parameters affect model performance, we tested different configurations on the learning rate, dropout rate, batch size, and the number of MLP layers. As shown in Fig.~\ref{fig:hyper_parameter_test}, the performance of the J2Q-TC-DAN model is mostly insensitive to its hyper-parameters besides the dropout rate, which is expected. Using a larger batch size slightly hurts the performance of our model due to overfitting~\cite{keskar_large-batch_2017}. Lastly, we found that using three-layer MLP works the best for our question template classification task.
\vspace{-.2cm}
\subsubsection{Question Ranking.} We evaluated four question ranking models listed in Sec.~\ref{sec:exp_setting} using $22,055$ (job, template, parameter) triples from $6,675$ jobs and report the Area Under the Receiver Operating Characteristic curve (AUROC), Precision@k, Recall@k, and Normalized Discounted Cumulative Gain at $k$ (NDCG@$k$). 
As shown in Tab.~\ref{tab:qr_offline_eval} the proposed J2Q-QR-XGB-pairwise model outperforms other baselines with up to $24.03\%$ improvement in NDCG. 
This significant improvement indicates the effectiveness of the proposed model on predicting job poster actions on screening questions. Based on the observation, we chose the pairwise J2Q model as our online question ranking model due to its great ranking performance. 

\begin{table}[t]
    \caption{Question ranking offline evaluation.}
    \vspace{-.2cm}
    \label{tab:qr_offline_eval}
    \centering
    \begin{adjustbox}{max width=\linewidth}
        \begin{tabular}{l|c|cccccc}
            \multirow{2}{*}{Model} & \multirow{2}{*}{AUROC} & \multicolumn{2}{c}{Precision} & \multicolumn{2}{c}{Recall} & \multicolumn{2}{c}{NDCG} \\
             &  & @1 & @3 & @1 & @3 & @1 & @3 \\\hline
            Rule-based & $0.6408$ & $0.6795$ & $0.4505$ & $0.5499$ & $0.8967$ & $0.6795$ & $0.8075$ \\
            J2Q-QR-LR & $0.8008$ & $0.8205$ & $0.4864$ & $0.6719$ & $0.9629$ & $0.8205$ & $0.9078$ \\
            J2Q-QR-XGB-Pointwise & $\mathbf{0.8282}$ & $0.8325$ & $0.4876$ & $0.6818$ & $0.9650$ & $0.8325$ & $0.9136$ \\
            J2Q-QR-XGB-Pairwise & $0.8164$ & $\mathbf{0.8428}$ & $\mathbf{0.4897}$ & $\mathbf{0.6910}$ & $\mathbf{0.9672}$ & $\mathbf{0.8428}$ & $\mathbf{0.9194}$
        \end{tabular}
    \end{adjustbox}
    \vspace{-.2cm}
\end{table}

\begin{table}[t]
    \caption{Question ranking feature ablation study.}
    \vspace{-.1cm}
    \label{tab:qr_feature_ablation_study}
    \centering
    \begin{adjustbox}{max width=\linewidth}
    \begin{tabular}{l|c|cccccc}
    \multirow{2}{*}{Model} & \multirow{2}{*}{AUROC} & \multicolumn{2}{c}{Precision} & \multicolumn{2}{c}{Recall} & \multicolumn{2}{c}{NDCG} \\
     & & @1 & @3 & @1 & @3 & @1 & @3 \\\hline
    J2Q-QR-XGB-Pairwise & $0.8164$ & $\mathbf{0.8428}$ & $\mathbf{0.4897}$ & $\mathbf{0.6910}$ & $\mathbf{0.9672}$ & $\mathbf{0.8428}$ & $\mathbf{0.9194}$ \\
    \quad\quad -- no job feat. & $\mathbf{0.8170}$ & $0.8374$ & $0.4889$ & $0.6861$ & $0.9661$ & $0.8374$ & $0.9167$ \\
    \quad\quad -- no question feat. & $0.8077$ & $0.8287$ & $0.4858$ & $0.6801$ & $0.9622$ & $0.8287$ & $0.9104$ \\
    \quad\quad -- no interaction feat. & $0.8148$ & $0.8414$ & $0.4879$ & $0.6889$ & $0.9650$ & $0.8414$ & $0.9168$
    \end{tabular}
    \end{adjustbox}
    \vspace{-.4cm}
\end{table}



\subsubsection{Ranking Feature Ablation Study.} In this experiment, we study which feature group contributes the most to the question ranking model. In Tab.~\ref{tab:qr_feature_ablation_study}, we reported the AUROC and NDCG@$k$ of different J2Q-QR-XGB-pairwise models trained with one group of the features removed. The results show that all feature groups positively contribute to the model and should be retained. Note that adding interaction features only improve the performance marginally. This is because tree-based model can capture some feature correlations without explicit signal.

\subsection{Online Evaluation}
\subsubsection{Screening Question Suggestions.} When posting jobs on LinkedIn, posters can manually add screening questions to jobs. Here we deployed two template classification mdoels, BOW-XGB and J2Q-TC-DAN, online to provide SQ suggestions to posters. We ramped each model to $50\%$ of LinkedIn's traffic for two weeks and compared them against a baseline model, which simply extracts parameters from job posting and create questions regardless the sentence intent. The metrics we tracked are as follows:

\begin{itemize}[leftmargin=*]
    \item \textbf{Acceptance Rate}: \#accepted SQs / \#suggested SQs,
    
    \item \textbf{Suggestion Rate}: \#jobs receive SQ suggestions / \#jobs on LinkedIn,
    
    \item \textbf{Adoption Rate}: \#jobs accepted SQ suggestions / \#jobs on LinkedIn.
\end{itemize}

As shown in Tab.~\ref{tab:ab_question_suggestions}, the proposed Job2Questions model significantly improves both precision and coverage. We believe the acceptance improvement is due to the expressive power of neural network models on modeling job posting semantics. The increase in coverage, on the other hand, is because the deep transfer learning gives good generalization ability to our model so it can handle job postings written in different styles.

\begin{table}[t]
    \caption{Screening question suggestion online A/B test.}
    \vspace{-.2cm}
    \label{tab:ab_question_suggestions}
    \begin{adjustbox}{max width=\linewidth}
        \begin{tabular}{l | c c c}
        Model & Acceptance Rate & Suggestion Rate & Adoption Rate\\
        \midrule
        BOW-XGB & $+31.06\%$ & $+40.86\%$ & $+14.26\%$\\
        Job2Questions & $+\mathbf{53.10}\%$ & $+\mathbf{59.92}\%$ & $+\mathbf{22.17\%}$ \\
        \end{tabular}
    \end{adjustbox}
\end{table}

\begin{figure}[t] 
\centering
\vspace{-.2cm}
  \begin{subfigure}[b]{\linewidth}
    \centering
    \includegraphics[width=.75\textwidth]{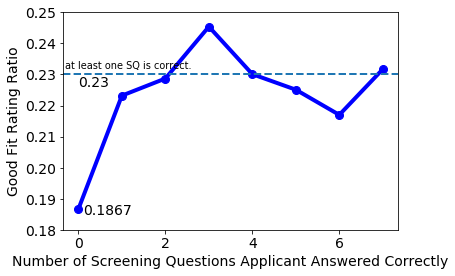}
  \end{subfigure}%
\vspace{-.2cm}
  \caption{Applicants' good fit rating versus the number of screening questions they answered correctly. The horizontal line denotes the average good fit rating ratio of applicants who answered at least one question correctly.}
\vspace{-.4cm}
\label{fig:good_fit_rating} 
\end{figure}
\vspace{-.2cm}
\subsubsection{Job Applicant Quality.} After the proposed Job2Questions model is deployed, we also analyzed how providing screening questions can help improve the hiring efficiency in terms of job applicant quality. 
We first look at how screening questions help identify qualified applicants in around $20$M job applications. As shown in Fig.~\ref{fig:good_fit_rating}, we found that if the applicant does not answer any screening questions correctly, only $18.67\%$ of the applications are rated as a good fit by the recruiter. But if the applicant answered at least one screening question correctly, the good fit rating increases to $23\%$ ($+23.19\%$).
\vspace{-.2cm}
\subsubsection{Hiring Efficiency.} Because recruiters often receive a large amount of applications per job, it is impractical for them to review all of them. Therefore they tend to sort the applicants first and only review the top-ranked candidates. Here at LinkedIn we sort the candidates by predicting if an applicant is a good fit. As an alternative, we developed another ranking model that simply sort applicants based on the favorableness of their answers to the screening questions. 
We conducted a $50$-$50$ A/B test for one month and measured the good/bad fit rating that recruiters gave to the candidates they contacted. We found that ranking applicants by screening question answers can improve the applicant good fit rate by $7.45\%$ and reduce the bad fit rate by $1.67\%$. This means the screening questions can help the recruiters surface qualified applicants and therefore improve the hiring efficiency.
\vspace{-.2cm}
\subsubsection{Screening Question-based Job Recommendation.} We also conducted applicant-side analysis to see if SQs can help applicants apply for jobs they are qualified for. We applied our Job2Questions model to all jobs on LinkedIn and retrained our job recommendation model (JYMBII~\cite{kenthapadi_personalized_2017}) using SQs and applicant answers as additional features. We observed that when LinkedIn members applying for jobs suggested via email, they are $46\%$ more likely to get a good fit rating if the job is suggested by the JYMBII + SQ model.

\vspace{-.2cm}
\subsubsection{Increased Interactions and Satisfactions.} By providing SQ suggestions to recruiters, we observed boosted positive interactions. Namely jobs with SQs yield $1.9$x more recruiter-applicant interactions in general and $2.4$x more interactions with screening-qualified applicants. Moreover, the Net Promoter Score (NPS)~\cite{reichheld2003one} is $11$ points higher for recruiters using SQs than those who don’t.

\subsection{Case Studies and Insights}\label{sec:case_studies}

\subsubsection{SQ Answers Complement Member Profile.} To verify our hypothesis that member profile is not an ideal data source for job-applicant fit measurement, we compared the member profile and their screening question answers. We found that screening questions often contains information that members do not put in their profile. In Tab.~\ref{tab:sq_profile}, we can see that among members who answered screening questions, $33\%$ of the members do not provide their education information in their profile. More specifically, people who hold secondary education degree are less likely to list that in their profile. As for languages, $70\%$ of the members do not list the languages they spoke (mostly native speakers) in their profile. Lastly, $37\%$ of the members do not include experience with specific tools, e.g. \textit{Salesforce Sales Cloud}, \textit{Adobe Design Programs}, or \textit{Google Ads}, in their profile. In short, we suspect that when people composing their professional profile, they tend to overlook basic qualifications which recruiters value a lot during screening. Therefore, screening questions are much better, direct signals for applicant screening compared to member profile.

\begin{table}[t]
    \caption{Relationships between SQ answers and user profile.}
    \vspace{-.2cm}
    \label{tab:sq_profile}
    \begin{adjustbox}{max width=\linewidth}
        \begin{tabular}{l|rrr}
        & Edu. & Lang. & Tools \\\hline
        Matches Profile (\%) & $64\%$ & $29\%$  & $61\%$ \\
        Not in Profile (\%) & $33\%$ & $70\%$  & $37\%$ \\
        Conflicts with Profile (\%) & $3\%$ & $1\%$ & $2\%$
        \end{tabular}
    \end{adjustbox}
    \vspace{-.3cm}
\end{table}


\begin{table}[t]
    \caption{Question rejection rate case study.}
    \vspace{-.2cm}
    \label{tab:question_rej_rate_case_study}
    \begin{adjustbox}{max width=\linewidth}
        \begin{tabular}{l|c|c}
            \multicolumn{1}{c|}{Question}                                           & Type & Rej. rate \\ \hline
            How many years of work experience do you have using \underline{Fax}?               & Tool & $99.10\%$   \\
            How many years of work experience do you have using \underline{Internet Explorer}? & Tool & $98.75\%$   \\
            Have you completed the following level of education: \underline{master's degree}?  & Edu. & $77.68\%$  
        \end{tabular}
    \end{adjustbox}
    \vspace{-.2cm}
\end{table}

\begin{table}[t]
    \caption{Per-industry SQ type distribution.}
    \vspace{-.2cm}
    \label{tab:question_acc_rate}
    \begin{adjustbox}{max width=\linewidth}
    \begin{tabular}{l|rrrrrr}
    Industry & Cred. & Edu. & Lang. & Sponsor & Tools & Work Auth. \\
    \toprule
    Agriculture & $3.46\%$ & $\mathbf{36.88}\%$ & $\mathbf{28.03}\%$ & $2.94\%$ & $21.47\%$ & $7.22\%$ \\
    Government & $8.99\%$ & $30.70\%$ & $15.72\%$ & $\mathbf{4.25}\%$ & $26.47\%$ & $13.86\%$ \\
    Technology & $0.79\%$ & $3.82\%$ & $2.31\%$ & $0.73\%$ & $\mathbf{90.81}\%$ & $1.54\%$ \\
    Transportation & $\mathbf{15.47}\%$ & $25.54\%$ & $19.95\%$ & $3.95\%$ & $20.94\%$ & $\mathbf{14.14}\%$\\
    \hline
    \textbf{Overall} & $4.96\%$ & $10.89\%$ & $6.39\%$ & $1.45\%$ & $72.47\%$ & $3.84\%$ 
    \end{tabular}
    \end{adjustbox}
    \vspace{-.2cm}
\end{table}

\vspace{-.2cm}
\subsubsection{Job Postings Are Noisy.} Because our Job2Questions model generate questions for explicitly mentioned requirements only, we can identify requirements that recruiters think trivial or unnecessary by finding suggested SQs with top rejection rate, i.e. requirements mentioned in the job posting but not important enough to be actual screening questions. Tab.~\ref{tab:question_rej_rate_case_study} shows top-$3$ SQs with the highest rejection rate. We found that although recruiters explicitly mention requirements such as ``\textit{Access to computer with scanning, printing and \textbf{faxing} capabilities}'' or ``\textit{Good working knowledge of \textbf{Internet Explorer}}'', more than $98\%$ of the cases recruiters do not screen applicants based on these. We suspect recruiters do not update job postings frequently, therefore it sometimes contain outdated requirements that are too trivial to be used for screening. Another interesting finding is that job postings often contain unnecessary requirements such as degree requirements. Although job postings explicitly state requirements such as ``\textit{A Bachelor of Science or a \textit{Master Degree} required}'', recruiters usually screen applicants based on the lowest education requirement only. Based on these observations, we believe SQs are better than noisy job descriptions for modeling job requirements because they are more concise and reflect only the true needs of jobs.
\vspace{-.2cm}
\subsubsection{SQ preferences across industries.} Lastly, we found that different industries has different preferences or focus on screening candidates. In Tab.~\ref{tab:question_acc_rate}, we presented the overall SQ types used by all jobs posted on LinkedIn and per-industry breakdown of four example industries with interesting trends. For example, Agriculture industry is $4.4$ times more likely to screening applicants based on language than other industries in general. Technology industry does not screening candidates based on education or language, instead $91\%$ of the SQs are about tools they have used. Transportation industry does not require tools experience but are more likely to screen candidates by credentials such as driver's license or license to handle hazardous materials. Government and Transportation both ask a lot of work authorization and sponsorship questions probably because they usually do not sponsor working visas for foreigners but they do get a lot applicants who need sponsorship. By looking at SQ type distributions, we can better understand what each industry is looking for and how applicants can better position themselves by excelling in things employers value the most.

\section{Conclusions and Future Work}
In this work, we proposed a novel Screening Question Generation (SQG) task that automatically generates screening questions for job postings. We also developed a general candidate-generation-and-ranking SQG framework and presented LinkedIn's in-production Job2Questions model. We provided design details of Job2Questions, including data preparation, deep transfer learning-based question template classification modeling, parameter extraction, and XGBoost-based question ranking. The extensive online and offline evaluations demonstrate the effectiveness of the Job2Questions model. 

As for future work, we plan to infer SQs that are not explicitly mentioned in the job posting and investigate advanced question ranking methods to better model recruiter preferences. \hl{We also plan to investigate seq2seq models for template-free SQ generation.}.

\clearpage
\bibliographystyle{ACM-Reference-Format}
\bibliography{acmart}

\end{document}